\documentclass[twocolumn,amsmath,showpacs]{revtex4}

\usepackage{graphicx}
\usepackage{color}
\usepackage{hyperref}
\begin{document}

\newcommand{\MVAL}{190}

\title{Observing microscopic structures of a relativistic object\\ using a “time-stretch” strategy}
\author{E. Roussel$^{1,2}$, C. Evain$^{1,2}$, M. Le Parquier$^2$,
  C. Szwaj$^{1,2}$, S. Bielawski$^{*1,2}$}
\affiliation{($1$) Laboratoire PhLAM, UMR CNRS 8523, Universit\'e Lille 1,
  Sciences et Technologies, 59655 Villeneuve d'Ascq, France\\
($2$) Centre d'\'Etude Recherches et Applications (CERLA), 59655 Villeneuve d'Ascq, France}
\author{ L. Manceron$^3$,
  J.-B. Brubach$^3$, M.-A. Tordeux$^3$, J.-P. Ricaud$^3$,
  L. Cassinari$^3$, M. Labat$^3$, M.-E Couprie$^3$, P. Roy$^3$}
\affiliation{($3$) Synchrotron SOLEIL, L'Orme des Merisiers, Saint-Aubin, BP 48, 91192
  Gif-sur-Yvette Cedex, France\\
 (*) serge.bielawski@univ-lille1.fr
}

\date{\today}
\maketitle

{\bf Emission of light by a single electron moving on a curved trajectory
  (synchrotron radiation) is one of the most well-known fundamental radiation
  phenomena. However experimental situations are more complex as they involve
  many electrons, each being exposed to the radiation of its neighbors. This
  interaction has dramatic consequences, one of the most spectacular being the
  spontaneous formation of spatial structures inside electrons bunches. This
  fundamental effect is actively studied as it represents one of the most
  fundamental limitations in electron accelerators, and at the same time a
  source of intense terahertz radiation (Coherent Synchrotron Radiation, or
  CSR).  Here we demonstrate the possibility to directly observe the electron
  bunch microstructures with subpicosecond resolution, in a storage ring
  accelerator. The principle is to monitor the terahertz pulses emitted by the
  structures, using a strategy from photonics, time-stretch, consisting in
  slowing-down the phenomena before recording. This opens the way to
  unpreceeded possibilities for analyzing and mastering new generation high
  power coherent synchrotron sources.}

{Interaction of a relativistic electron bunch with its own created
  electromagnetic field can lead to the so-called {\it microbunching
    instability}. {It is encountered in systems based on linear
    accelerators~\cite{PhysRevSTAB.10.104401}, solar
    flares~\cite{kauffmann_solarflares_phys_plasma13_070701_2006,kauffmann_solarflares_solarphys255_131_2009,kauffmann_solarflares_astrophysicaljournal_791_2014},
    as well as in the widely-used storage rings
    facilities~\cite{stupakov2002.PhysRevSTAB.5.054402,byrd2002.PhysRevLett.89.224801,abo2002.PhysRevLett.88.254801,hashimoto2005.1591760,mochihashi2006,ANKA.IPAC12.TUPP010,DIAMOND_fitRL,MLS.IPAC10,Karantzoulis2010300,evain2012.0295-5075-98-4-40006}
    (synchrotron radiation facilities), where electron bunches are forced to
    circulate onto a closed loop trajectory}. Above a threshold electron bunch
  density, a longitudinal modulation or {\it pattern} appears with a
  characteristic period at the millimeter or sub-millimeter
  scale~\cite{stupakov2002.PhysRevSTAB.5.054402,byrd2002.PhysRevLett.89.224801,abo2002.PhysRevLett.88.254801,Warnock:2006qa}.}
This structure emits intense pulses of terahertz radiation (typically more
than 10000 times normal synchrotron radiation), called Coherent Synchrotron
Radiation (CSR). Each CSR pulse shape may be viewed as an ``image'' of the
electron bunch microstructure.

\begin{figure}[htbp]
  \centering
  \includegraphics[width=8.5cm]{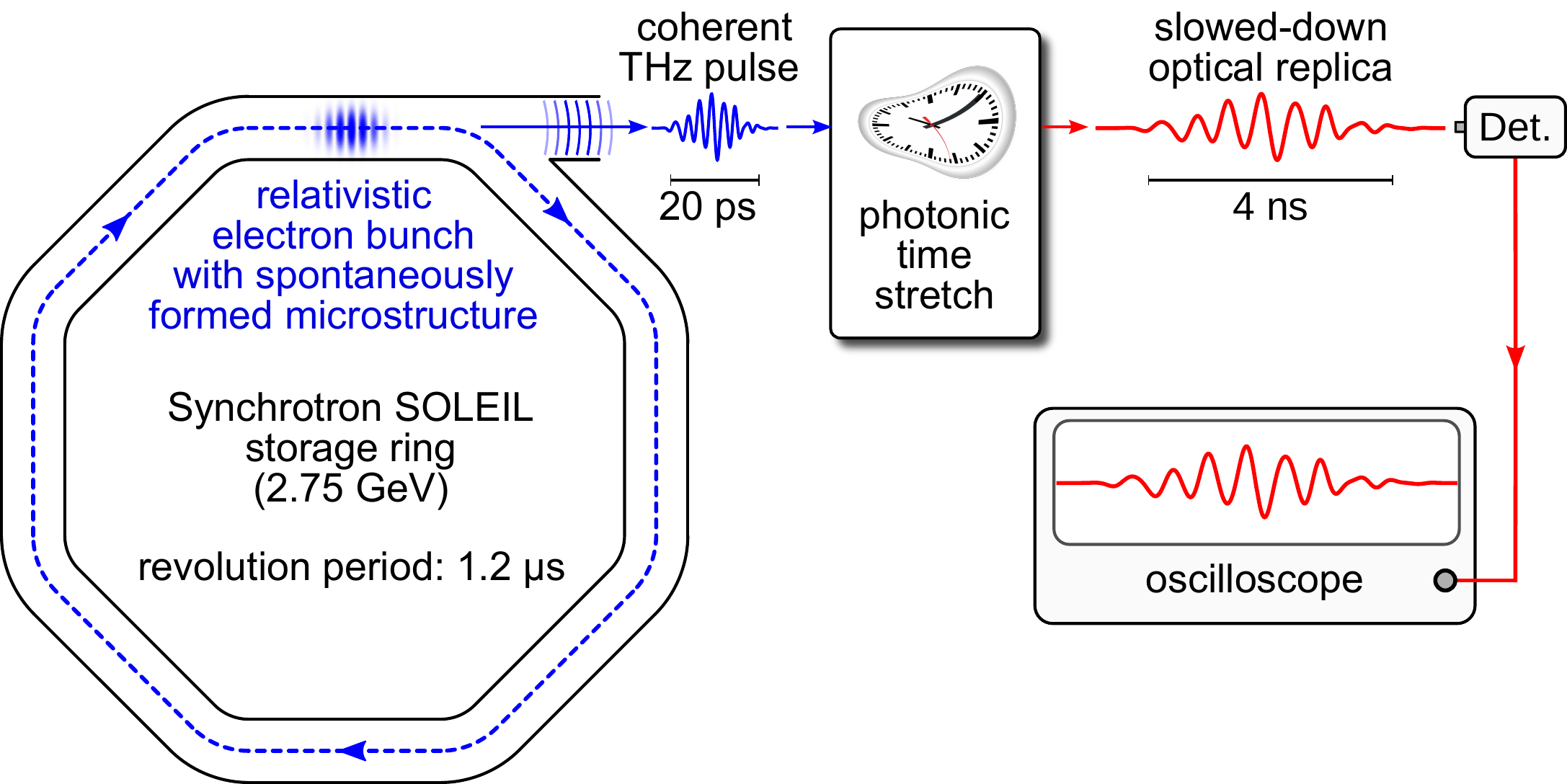}
  \caption{General principle of the experiment. A relativistic electron bunch
    circulating in the SOLEIL storage ring (2.75~GeV) presents
    microstructures, that evolve in a complex way. The coherent THz radiation
     emitted at a bending magnet carries the information on the
    microstructure shape, but is too fast (at picosecond scale) to be recorded
    by traditional means. The present strategy consists in ``slowing-down'' the
    information in order to obtain an optical replica at the nanosecond scale,
    so that a conventional oscilloscope can be used for the recording. }
  \label{fig:global_setup}
\end{figure}

As a consequence, a particularly efficient way to study this
fundamental physical effect consists in using existing
{\it user-oriented} storage rings (i.e., synchrotron radiation
facilities). Indeed, recording CSR pulses emitted at each turn in such
a storage ring would be theoretically sufficient to follow the electron
microstructure evolution over a long time. This has been recently
demonstrated in a special case where the microstructure wavelength is
in the centimeter range and CSR emission occurs in the tens of GHz
range~\cite{uvsor_MBI_ybco}, thus being accessible to conventional
electronics. However, in most storage rings such as
ALS~\cite{byrd2002.PhysRevLett.89.224801},
ANKA~\cite{ANKA.IPAC12.TUPP010},
BESSY~\cite{abo2002.PhysRevLett.88.254801},
DIAMOND~\cite{DIAMOND_fitRL,MLS.IPAC10}, ELETTRA~\cite{DIAMOND_fitRL},
or SOLEIL~\cite{evain2012.0295-5075-98-4-40006}, etc., {the CSR
  emission occurs at frequencies that are so high (above 300~GHz for
  the present case of SOLEIL) that no suitable recording electronics
  is available at the moment, nor expected in the near future.}

\begin{figure*}[htp]
  \centering
  \includegraphics[width=17cm]{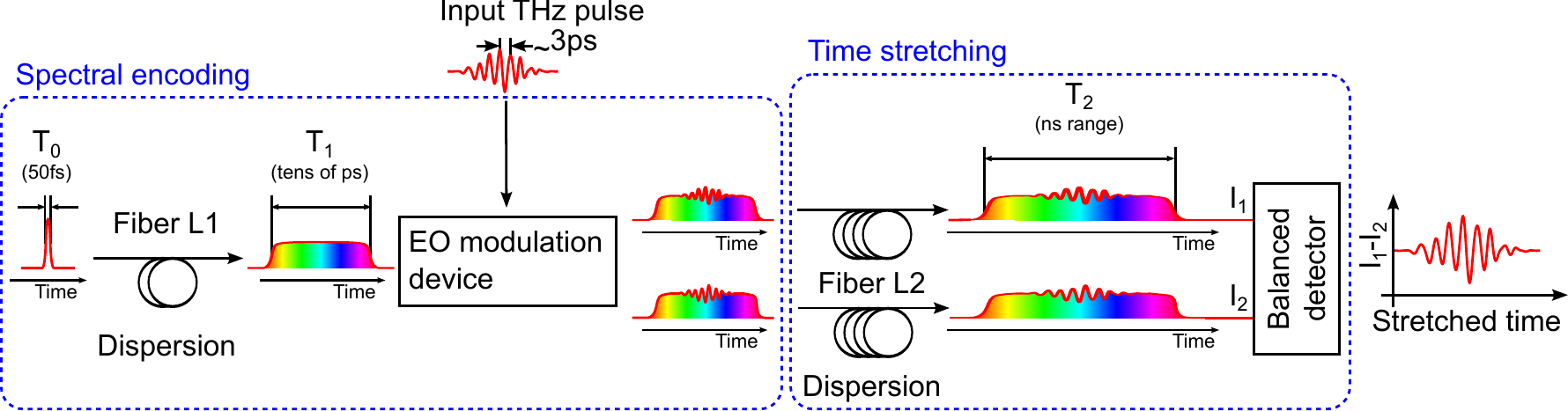}
  \caption{Principle of the photonic time-stretch device realized for slowing
    down the information, while keeping a high sensitivity. The THz pulse
    under investigation is first encoded into a chirped laser pulse, using an
    electro-optic crystal (``EO modulation device''). This device (see
    Fig.~\ref{fig:detailed_EOS_setup} for detils) provides two complementary
    outputs. Then the optical information of each output is simply stretched
    from picoseconds to nanoseconds by propagation in a long fiber
    (2~km). Balanced detection is performed between the two stretched laser
    pulses, thus providing a very high sensitivity for the device by removing
    the ``DC'' background. Details of the optical system are presented in the
    Methods section.}
  \label{fig:principle_time_stretch}
\end{figure*}

Here, we propose a strategy that overcomes these limitations, and thus enables
such fundamental studies in many storage ring facilities. It consists in
``slowing down'' the signals so that they can be recorded by conventional
oscilloscopes (Fig.~\ref{fig:global_setup}). This is a two-step process as
shown in Figure~\ref{fig:principle_time_stretch}. First, the THz CSR pulse is
encoded into a laser pulse, using the well established technique of THz
electro-optic
sampling~\cite{EOS_first_spectral_encoding,PhysRevLett.88.124801,PhysRevSTAB.15.070701,ANKA.IPAC12.TUPP010}. Then
the key point is to use a setup {based on} the so-called {\it photonic
  time-stretch}
strategy~\cite{time_stretch_first_Coppinger_1999,PhysRevA.45.600,jalali_nature_photonics_review},
which consists in dispersing the pulses in a long fiber. Under some condition
on the fiber length $L_1$, the output pulse is a replica of the original
signal, except that is slowed down by a magnification factor (or stretch
factor)~\cite{time_stretch_transfer_function_han_2003}:
\begin{equation} 
M=1+\frac{L_2}{L_1},
\end{equation}
where $L_1$ and $L_2$ are the input and output fiber {lengths} ($L_1=10.7$~m
and $L_2=$2~km for the results presented hereafter, leading to a stretch
factor $M=\MVAL$). Using such a strategy instead of the classical {\it
  spectral
  encoding}~\cite{EOS_first_spectral_encoding,PhysRevLett.88.124801,PhysRevSTAB.15.070701}
method presents two advantages: (i) {a much} higher acquisition rate (i.e.,
the number of recorded terahertz pulses per second), which is directly linked
to the laser repetition rate, and (ii) the possibility of performing balanced
detection~\cite{wong2011_time_stretch_balanced}, a crucial point for reaching
high sensitivity. 

{The time-stretch THz recording system (detailed in
   Figure~\ref{fig:detailed_EOS_setup}
) was able to acquire 88~$\times 10^6$
  terahertz pulses per second, and its sensitivity was measured to be
  37~V/cm inside the crystal (see Methods). This allowed us to record the
  terahertz CSR pulses (electric field, including envelope and
  carrier) emitted at each turn (i.e., every 1.2 ~$\mu$s) at the AILES
  beamline of the SOLEIL storage ring.}

\begin{figure}[htbp] 
  \centering
  \includegraphics[width=8.5cm]{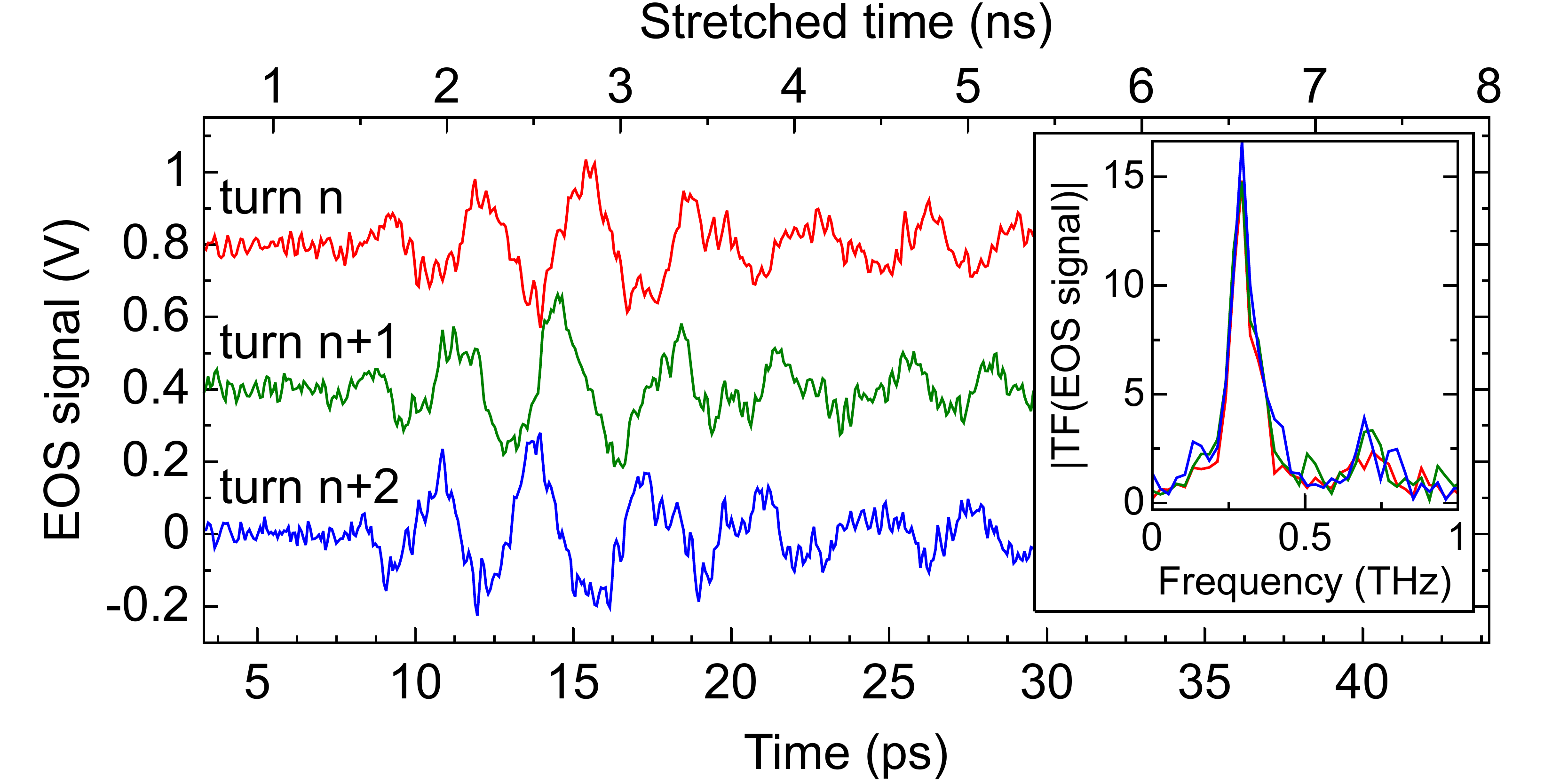}
  \caption{Typical single-shot recordings of coherent THz pulses,
    which carry the information on the electron microstructure. The
    lower scale corresponds to the real time of the phenomenon. The
    upper scale corresponds to the time at the oscilloscope input
    (i.e., after the photonic time stretch by a factor \MVAL). The time
    between pulses is of the order of 1~$\mu$s, but the system is
    actually recording one signal every 12~ns. The effective A/D
    conversion speed is 15~Tera samples/s. Inset: Power Fourier
    spectra of the three pulses.}
  \label{fig:exp_traces}
\end{figure}

{First experiments with the time-stretch acquitition
  setup allowed us to record the THz CSR for each revolution in the
  ring, and to visualize the predicted microstructure in the electron
  bunch circulating at Synchrotron SOLEIL. The structure is clearly
  visible in Figure~\ref{fig:exp_traces}, which represents a typical
  series of individual pulses, recorded at successive round-trips.}

\begin{figure}[htbp]
  \centering
  \includegraphics[width=8.5cm]{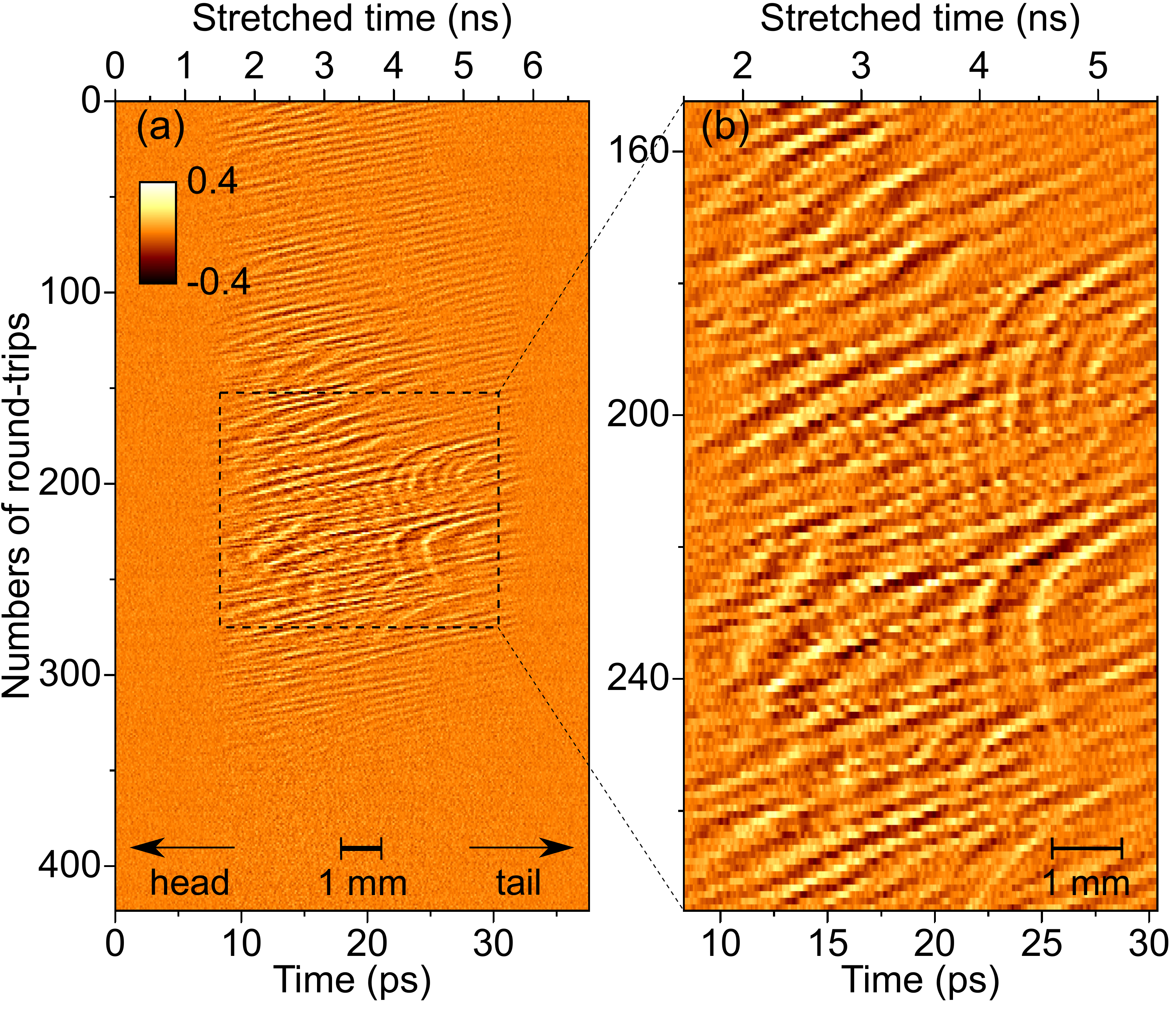}
  \caption{Recorded CSR pulses (i.e., containing the information
    on the electron bunch longitudinal modulation) versus the number
    of round-trips in the storage ring. 
    (b) is a zoom on the rectangle part in (a).}
  \label{fig:exp_traces_colorscale}
\end{figure}

{Furthermore, the new type data thus obtained contain
  extremely detailed information on the long-term evolution of the
  structures. In order to summarize the dynamical features, we
  displayed the pulse evolution of Fig.~\ref{fig:exp_traces} as a
  colormap versus the longitudinal coordinate, and the number of turns
  in the storage ring~(Fig.~\ref{fig:exp_traces_colorscale}). It
  clearly appears that --though their evolutions are very complex--
  the structures are constituted of oscillations with a characteristic
  wavenumber in the millimeter range (more precisely 10~cm$^{-1}$, or
  0.3~THz as shown in the inset of Fig.~\ref{fig:exp_traces}). This is
  consistent with previous indirect observations, by spectrometric
  measurements, of a strong terahertz emission peak at
  10~cm$^{-1}$~\cite{evain2012.0295-5075-98-4-40006}. Moreover, the
  recordings (as in Fig.~\ref{fig:exp_traces_colorscale})
  systematically revealed complex drifting evolution during the
  revolutions, that remind the ubiquitous irregular behaviors that
  occur in fluid dynamics~\cite{cross1993.RevModPhys.65.851}. We
  believe that this new detailed data will provide a real platform for
  testing and refining theoretical models of relativistic electron
  bunch dynamics.}

{Several important features of the experimental
  observations can be reproduced with already existing theoretical
  models~\cite{stupakov2002.PhysRevSTAB.5.054402,venturini2002.PhysRevLett.89.224802,evain2012.0295-5075-98-4-40006,roussel2014.PRSTAB.17.010701.2014},
  where only longitudinal dynamics of the electrons is taken into
  account. Each electron $i$ is characterized by its instantaneous
  position $z_i$ and energy variable $\delta_i$. A map (at each turn
  $n$) can then be written for the evolutions of $z_i^n$ and
  $\delta_i^n$. Taking the continuous limit for the number of
  round-trips~$n$~\cite{PhysRevE.55.3493,ChaoBook_collective}:
\begin{eqnarray}
  \frac{dz_i}{dt}&=&-\eta c\delta_i\label{eq:model_mp_1}\\
  \frac{d\delta_i}{dt}&=&-\left(\frac{\omega_s^2}{\eta c}\right)z_i- \left(\frac{2}{\tau_s}\right)\delta_i-F(z_i)-\eta_{N}\xi(t) \label{eq:model_mp_2},
\end{eqnarray}
where $t$ is a continuous time associated to the number of round-trips
$n$.  $\delta_i(t)=[E_i(t)-E_R]/E_R$, with $E_i(t)$ the electron
energy, and $E_R$ the reference energy corresponding to the
synchronous electron (2.75~GeV here). $F(z_i)$ characterizes the
electric field at position $z_i$ created by the whole electron bunch,
and is the main ingredient of the instability. We use here the same
form as in previous studies of SOLEIL 
\cite{evain2012.0295-5075-98-4-40006}. $\omega_s/2\pi$ is the
synchrotron frequency (not to be confused with the storage ring
revolution time), and $\tau_s$ is the synchrotron damping time. $c$ is
the speed of light in vacuum and $\eta$ measures the dependence of the
round-trip time with the electron energy. $\eta_N\xi$ is a gaussian
white noise term, with $<\xi(t)\xi(t')>=\delta(t-t')$. Parameter values are given in the Methods section.}

\begin{figure}[htbp]
  \centering
  \includegraphics[width=8.5cm]{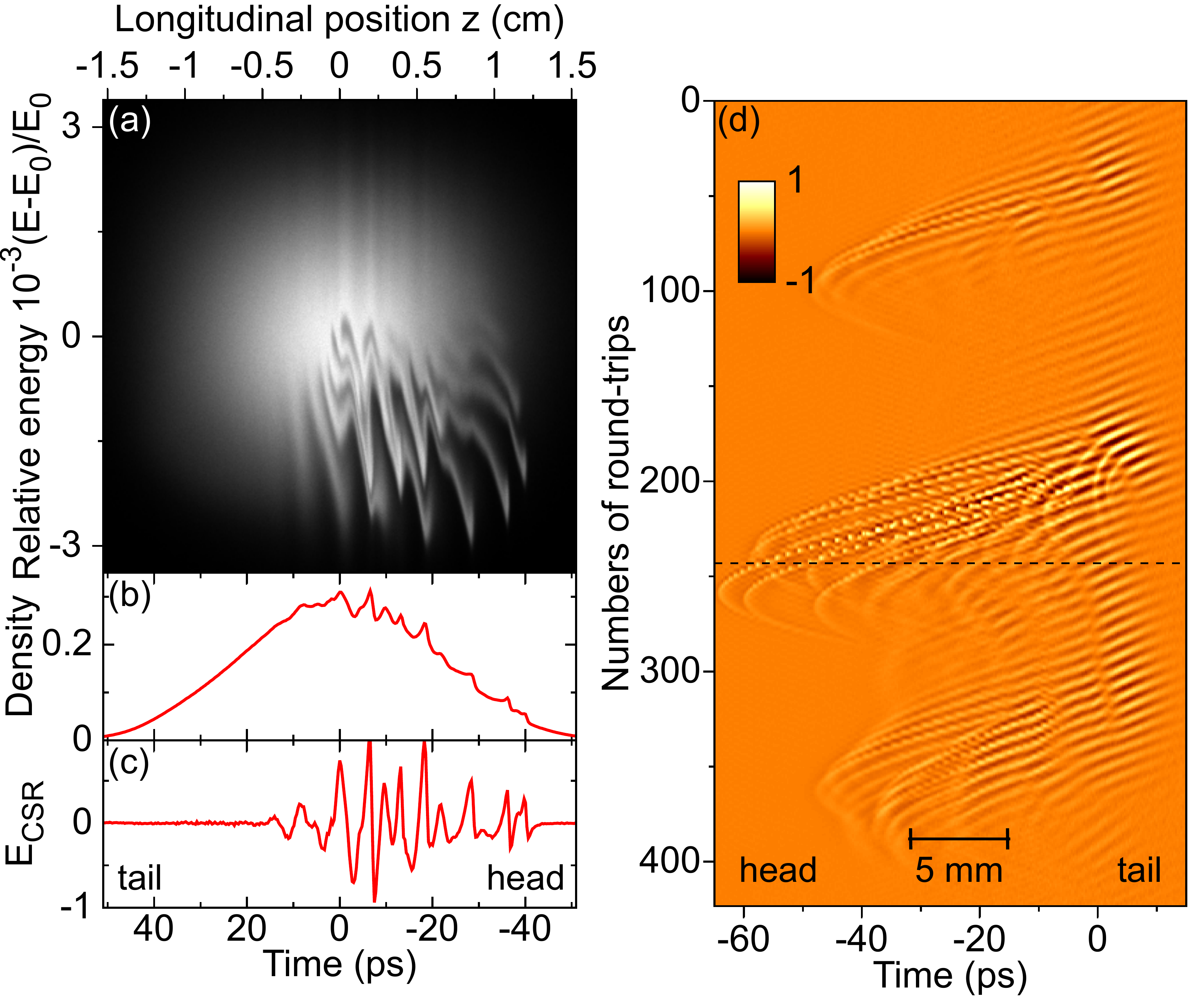}
  \caption{Numerical simulation of the relativistic electron bunch
    dynamics. (a) Distribution of the electrons in position-energy phase space
    ($1.1\times 10^8$ particles are taken into account in the simulation). The
    pattern globally rotates at a slow frequency near 4.64~kHz, and evolves in
    a complex way (see supplementary movie). (b) Electron bunch profile (i.e.,
    the projection of (a) onto the $z$ axis), and (c) CSR field. (d) evolution
    of the CSR pulse shapes versus number of round-trips. The
    dashed line corresponds to the (c) pulse.}
  \label{fig:num_traces_colorscale}
\end{figure}

Typical numerical results are presented in
Fig.~\ref{fig:num_traces_colorscale}. When the electron bunch charge exceeds a
threshold, finger-like structures are spontaneously formed in the electrons
phase space~(Fig.~\ref{fig:num_traces_colorscale}a). Furthermore, the whole
electron bunch distribution experiences a global rotation near the so-called
synchrotron frequency (4.64~kHz here), and evolves in a bursting and irregular
way.
 
The longitudinal electron bunch shape~(Fig.~\ref{fig:num_traces_colorscale}b)
is deduced from the vertical projection of the phase space distribution
(Fig.~\ref{fig:num_traces_colorscale}a). Then the CSR THz field at the
electron bunch location~(Fig.~\ref{fig:num_traces_colorscale}c) is deduced
from the electron bunch shape~(Fig.~\ref{fig:num_traces_colorscale}b). As can
be seen in Fig.~\ref{fig:num_traces_colorscale}c, only the fast variations
lead to a significant coherent terahertz field. This natural ``AC-coupling''
is an advantage for the observation as it removes the global (slow) electron
bunch shape, and let pass only the important information. Because the electron
bunch distribution rotates counter-clockwise in phase-space~(see
Fig.~\ref{fig:num_traces_colorscale}a and supplementary
movie), the microstructures drift along the longitudinal position toward
the head of the electron bunch (Fig.~\ref{fig:num_traces_colorscale}c). The
drift of the structures (in Figs.~\ref{fig:exp_traces_colorscale} and
~\ref{fig:num_traces_colorscale}d) can thus be interpreted as a consequence of
the formation of ``fingers'' in the lower part of phase space
(Figs.~\ref{fig:num_traces_colorscale}a). Main features of the theoretical
predictions are found in relatively satisfying agreement with the new
experimental findings.

In conclusion, the present time-stretch strategy allows one to perform a
``time-lapse observation'' of microscopic structures that appear within
charged relavististic objects. The advantages over classic single-shot
electro-optic sampling strategies are a simultaneous improvement on both the
aquisition rate and the sensitivity. {Such quantitative
  studies open-up to a new level of understanding of electron beam dynamics,
  and severe tests of theoretical models. We believe that this strategy may be
  a key contribution in situations where high acquisition rate measurements
  are needed. Straightforward applications concern the investigation of the
  THz pulses emitted by other storage rings, and by terahertz free-electron
  lasers. The technique can also be transfered to high-repetition rate linear
  accelerators, provided an electro-optic sampling system can be
  used~\cite{PhysRevLett.88.124801}. Perspectives in ultrafast spectroscopy
  are also envisaged, as the instantaneous THz spectrum can be
  straigtforwardly deduced from the electric field shape~(inset of
  Fig.~\ref{fig:exp_traces}). In addition to the THz domain, the present
  time-stretch strategy also opens new possibilities at short
  wavelengths. Important perspectives to be explored concern the monitoring of
  optical pulses from high-repetition rate EUV and X-ray Free-Electron Lasers,
  for instance by associating the time-stretch strategy to transient
  reflectivity setups~\cite{X_cross_co_LCLS,X_cross_co_FLASH_riedel_2013}.}

\section*{Methods}
\subsection{Ultrafast recording setup}
\begin{figure}
\includegraphics[width=8.5cm]{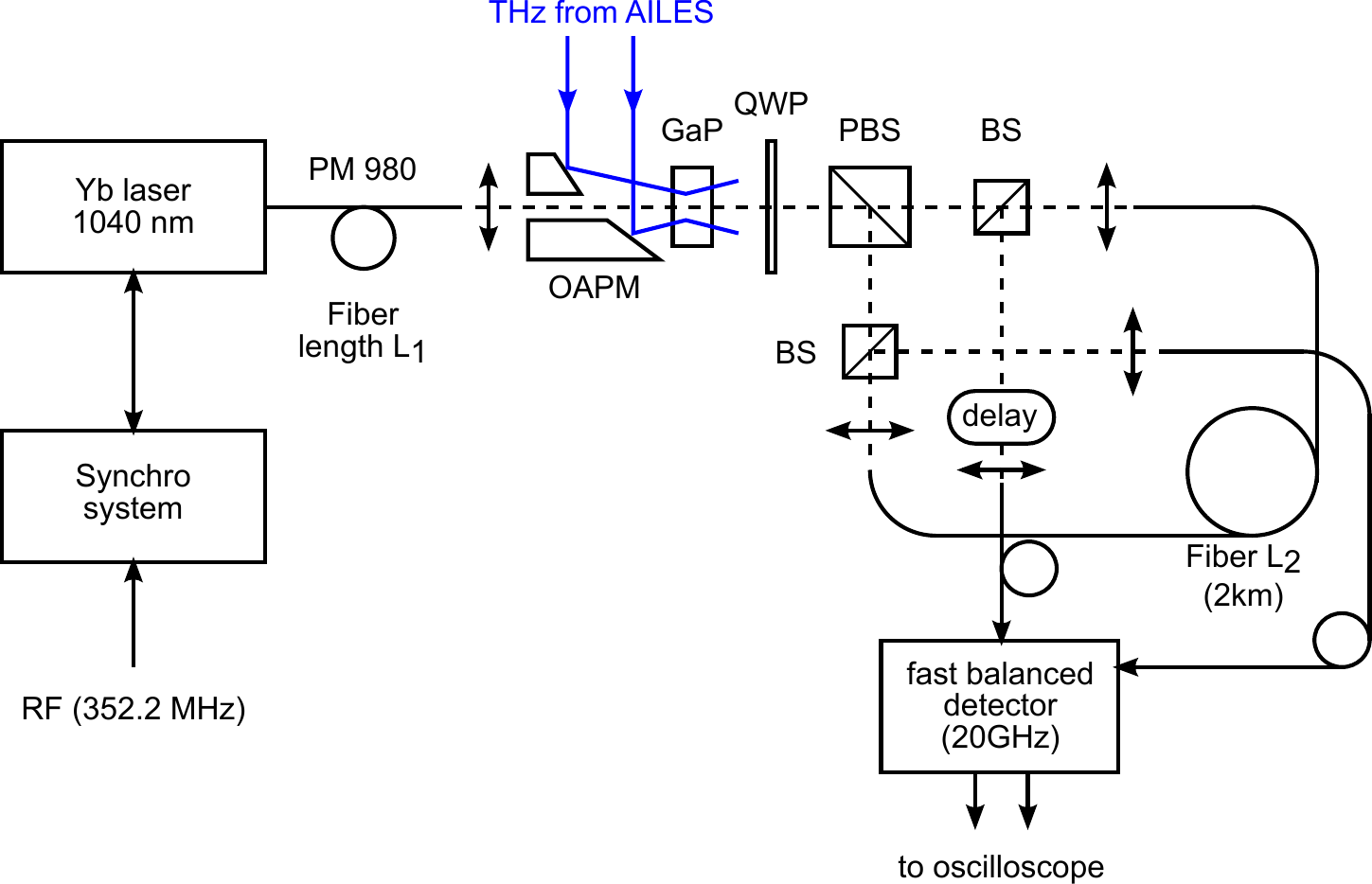}
\caption{Experimental setup for recording the THz
  pulses at 88~MHz acquisition rate. P: polarizer, OAPM: off-axis
  (gold-coated) parabolic mirror. GaP: Gallium Phosphide crystal. QWP:
  achromatic quarter-wave plate. PBS: polarizing beam-splitter. BS:
  beam-splitter with low polarization-dependent losses. The 2~km fiber is an
  HI1060 from Corning. All fiber collimating and focusing lenses have $11$~mm
  focal length. (Delay) is a delay line, allowing to adjust the relative delay
  between the two balanced detector inputs.}
\label{fig:detailed_EOS_setup}
\end{figure}

The detailed setup is presented in Fig.~\ref{fig:detailed_EOS_setup}.  It
exclusively uses off-the shelf components, and is composed of
three parts:
\begin{itemize}
\item A classical system for generation of chirped laser pulses, using a
  femtosecond laser and a dispersive fiber.
\item A classical electo-optic modulation system, based on the Pockels effect
  in a GaP crystal.
\item A specially designed balanced time-stretch device. This setup disperses
  the optical pulses up to the nanosecond range. Thus we can achieve
  simultaneously a high repetition rate, and at the same time a high
  sensitivity thanks to the possibility of the balanced detection. This is the
  key point of the setup.
\end{itemize}

\subsubsection{Production of the stretched laser pulses}
We use a femtosecond Ytterbium-doped fiber laser (Orange) from MENLO GmbH. The
emitted pulses have a spectral bandwidth of 50~nm, and the total output
average power is 40~mW. The repetition rate is chosen at 88~MHz, which
corresponds to 1/4~$th$ of the RF frequency of Synchrotron SOLEIL and
104~times the electron revolution frequency. It is synchronized on the RF
clock of the storage ring using a RRE-Synchro system from MENLO GmbH.

The pulses are chirped by a polarization-maintaining fiber (PM~980), which
length determines the temporal window of the acquisition (typically few tens
of picoseconds). The length $L_1$ which is used in the calculation of the
stretch factor $M=1+L_2/L_1$ is the sum of two components: the actual length
of the fiber used after the laser ($10$~m here), and a small contribution due
to pulse dispersion inside the laser (estimated to $0.7$~m of propagation in a
PM980 fiber). Thus we took $L_1=10.7$~m, leading to $M\approx \MVAL$.

A polarizer is placed at the output in order to remove possible spurious
components with the wrong polarization.

\subsubsection{Electro-optic modulation device}
This part corresponds to the {\it EO modulation device} in
Fig.\ref{fig:principle_time_stretch} and from (P) to (PBS) in
Fig.~\ref{fig:detailed_EOS_setup}. The THz radiation available at the focusing
point of the beamline is first collimated using an off-axis parabolic mirror
(not shown in Fig.~\ref{fig:detailed_EOS_setup}), with 101.6~mm focal
length. It is then focused in the GaP crystal using an off-axis gold-coated
parabolic mirror (OAPM2 in S) with 50.8~mm focal length, and a 3~mm hole. The
laser and the THz radiation interact in a [110]-cut GaP crystal with 5~mm
length (10x10x5 mm$^3$). The [-110] axis is parallel to the polarizations of
the laser and the THz beam. An achromatic quarter-wave plate is inserted after
the GaP crystal. Its optical axis oriented at $\pi/4$ with respect to the
laser polarization. Finally, a polarizing cube beam-splitter (PBS) provides
the two outputs of the EO modulation device. At the outputs of the PBS, the
pulses contain an intensity modulation which is a ``replica'' of the THz
pulses, and the two outputs are modulated in opposite phase.

\subsubsection{Time-stretching of the two optical pulses, using a single fiber}
Instead of using physically different fibers for the final dispersion (Fig.~\ref{fig:principle_time_stretch}), we use a variant that is much more robust from the
experimental point of view. As can be seen in
Fig.~\ref{fig:detailed_EOS_setup}, the two output pulses of the polarizing
cube beam-splitter are sent in the same fiber, in opposite
directions. Finally, two beam-splitters extract the two pulses, which are sent
to a fast balanced photoreceiver. This technique (which reminds the idea of
the Sagnac loop) permits to obtain the same path on the two laser pulses even
when the fiber optical length fluctuates.

 The $L_2$ fiber is an HI1060 from Corning with 2~km length (and an overall
 attenuation of the order of 3~dB). and the beam-splitters (BS) are chosen to
 have low polarization-dependent losses (Newport 05BC17MB.2). This choice for
 $L_2$ leads to stretched pulses of $\approx 4.5$~ns, and 1.35~mW peak power
 is typically dectected in each balanced photoreceiver channel input.

\subsection{Recording electronics}
The detection and subtraction between the two stretched signals is performed
using a DSC-R412 InGaAs amplified balanced photodetector from Discovery
Semiconductors, with 20~GHz bandwidth and 2800~V/W gain (specified at
1500~nm). The two differential outputs of the detector are sent on a Lecroy
LabMaster 10i oscilloscope with 36~GHz bandwidth, 80~GS/s sample rate on each
channel, and a memory of 256~Mega samples.

\subsubsection{Data processing}
In absence of THz signal, the recorded balanced signal presents a non-zero
shape which corresponds to imperfections of the setup, in particular small
polarization dependent losses (that depend on wavelength). Since this
``background'' signal is deterministic (i.e., is the same for each laser
pulse), it is easly removed from the signal, by a simple subtraction.

\subsubsection{Transport of the terahertz beam}
{We operated the time-stretch setup at the A branch of the AILES beamline,
  just before the interferometer~(see Ref.~\cite{roy.IPT.49.139.2006} for the
  beamlline detail). The focusing point was imaged onto the GaP crystal, using
  a telescope composed of a 101.6~mm focal length off-axis parabolic mirror
  (not shown in Fig.~\ref{fig:detailed_EOS_setup}) and a 50.8~mm off-axis
  parabolic mirror (OAPM2 in Fig.~\ref{fig:detailed_EOS_setup}).}

\subsection{Perfomances of the setup}
The special setup presented in Fig.~\ref{fig:detailed_EOS_setup} provides higher
acquisition rates (in terms of number of pulses per second) than classical
spectral encoding methods. The reason is that oscilloscopes can nowadays reach
much higher data acquisition rates (80 Giga samples/s here) than the cameras
which are necessary for spectral encoding. Here, the acquisition rate
capability is limited by the repetititon rate of the laser, namely 88~MHz.

At the same time, the setup allows us to perform acquisition with higher
sensitivity than the traditional single-shot electro-optic sampling, because,
of the possibilty to achieve balanced detection at the analog level. The
sensitivity is here mainly limited by the noise of the amplified balanced
detector. The RMS noise on the finally recorded signal can be easily measured,
and this gives a measure of the system sensitivity. The RMS noise (over the
20~GHz bandwidth of the phototetector) corresponds to a birefringence-induced
phase shift in the GaP of $3.2\times 10^{-3}$~Radian. Assuming that the
relevant electro-optic coefficient of GaP is $r_{41}=0.97$~pm/V, and
neglecting the THz frequency dependence of $r_{41}$, the sensitivity is
estimated to be $37$~V/cm (inside the crystal) near the laser pulse peak.


The SOLEIL storage ring was operated in single bunch, normal alpha
mode, with natural bunch length $\sigma_z=4.59$~mm, relative energy
spread $\sigma_\delta=1.017\times 10^{-3}$, and a momentum compaction
factor $\alpha=4.38\times 10^{-4}$. The ring was operated at a current
of $15$~mA, which is above the microbunching instability threshold
($\approx 10$~mA). Other parameters are described in
Ref.~\cite{evain2012.0295-5075-98-4-40006}. The model parameters
$\eta$ and $\eta_N$ are defined by $\alpha-\frac{1}{\gamma^2}$ and
$\eta_N=\frac{du_2}{T_0}$, with $T_0$ the storage ring revolution
time.


\section{Acknowledgements}
The work was supported by the Agence Nationale de la Recherche (projet Blanc
DYNACO 2010-042301), the Universit\'e Lille 1 (BQR 2012) and used HPC
resources from GENCI TGCC/IDRIS (2013-x2013057057, 2014x2014057057). The CERLA
is supported by the French Minist\`ere charg\'e de la Recherche, the R\'egion
Nord-Pas de Calais and FEDER.

\section*{Contributions}
{The acquisition system was designed, realized and
  operated by CS, ER, CE, MLP, SB, data analysis by CE and CS. Special
  operation of the AILES beamline was performed by LM, JBB, and PR
  (also project manager on SOLEIL side). Accelerator conditions were
  managed by MAT, diagnostics by ML, laser synchronization electronics
  by JPR, LC, MEC. MEC also participated to the early stages of the
  project elaboration. Numerical simulations were performed by ER,
  simulation code development by ER and SB.}

\bibliographystyle{./naturemag}


\end{document}